\global\def\draftcontrol{0}

   \def\versionno{ quiverthermo -- draft -- 4.05.03   }

\catcode`\@=11

\expandafter\ifx\csname
draftcontrol\endcsname\relax\global\def\draftcontrol{0} \fi

{\count255=\time\divide\count255 by 60
\xdef\hourmin{\number\count255} \multiply\count255
by-60\advance\count255 by\time
\xdef\hourmin{\hourmin:\ifnum\count255<10 0\fi\the\count255}}
\def\draftdate{\number\month/\number\day/\number\year\ \ \ \hourmin }

\newcommand\makepapertitle{\par
  \begingroup
    \renewcommand\thefootnote{\@fnsymbol\c@footnote}%
    \def\@makefnmark{\rlap{\@textsuperscript{\normalfont\@thefnmark}}}%
    \long\def\@makefntext##1{\parindent 1em\noindent
            \hb@xt@1.8em{%
                \hss\@textsuperscript{\normalfont\@thefnmark}}##1}%
     \newpage
     \global\@topnum\z@   
     \@makepapertitle
     \thispagestyle{empty}\@thanks
  \endgroup
  \setcounter{footnote}{0}%
  \global\let\thanks\relax
  \global\let\makepapertitle\relax
  \global\let\@makepapertitle\relax
  \global\let\@thanks\@empty
  \global\let\@author\@empty
  \global\let\@date\@empty
  \global\let\@title\@empty
  \global\let\title\relax
  \global\let\author\relax
  \global\let\date\relax
  \global\let\and\relax
  \def\version{\let\version\@version\@gobble}
}
\def\@makepapertitle{%
  \newpage
   \ifnum\draftcontrol=1 {}
   \version\versionno
   \vskip 5em%
   \else
   \hfill\hbox to 3cm {\parbox{4cm}{\@pubnum}\hss}%
   \vskip 5em%
   \fi
   \begin{center}%
   \let \footnote \thanks
     {
        {\Large\bf{\@title}}}%
     \vskip 1.5em%
     {\normalsize
       \lineskip .5em%
       \begin{tabular}[t]{c}%
         \@author
       \end{tabular}\par}%
     \vskip 1.5em%
     {\@bstract}%
     \end{center}%
     \vfill
     \@date%
     \vskip 1em
   \par
}

\gdef\@pubnum{}
\def\pubnum#1{%
  \gdef\@pubnum{#1}}

\gdef\@bstract{}
\def\Abstract#1{%
  \gdef\@bstract{%
   \parbox{\textwidth-0pc}{%
   \centerline{\bf Abstract}\penalty1000
   \noindent
   \renewcommand\baselinestretch{1.0}
   {#1}}}
}

\def\ps@paper{\let\@mkboth\@gobbletwo%
     \ifnum\draftcontrol=1
        \def\@oddfoot{\hbox to \textwidth{\tiny \versionno
\hfil\tiny\draftdate}%
        \hskip -\textwidth \hbox to \textwidth{\hfil\rm\thepage\hfil}}%
     \else\def\@oddfoot{\hbox to \textwidth{\hfil\rm\thepage\hfil}}
     \fi
     \let\@evenfoot\@oddfoot
}

\def\body{\clearpage
          \pagestyle{paper}
        }
\newenvironment{acknowledgements}{%
\vskip 3.25ex
\noindent {\bf Acknowledgements}
}

\def\@version#1{\ifnum\draftcontrol=1
\typeout{}\typeout{#1}\typeout{}
\vskip3mm\centerline{\hbox{\fbox{\normalsize{\tt DRAFT -- #1 -- }
                   {\draftdate}}}}\vskip3mm
\fi}
\let\version\@version
\long\def\eqlabel#1{\ifnum\draftcontrol=1
                    \tag@false  
                    \tag*{(\theequation) \hbox to
-0.2cm{\hspace{0cm}\small{#1}\hss}}
                    \refstepcounter{equation}
                    \edef\@currentlabel{\theequation}
                    \ltx@label{#1}          
                    \else
                    \label{#1}
                    \fi
                    }
\let\st@bibitem\@bibitem
\let\st@lbibitem\@lbibitem
\ifnum\draftcontrol=1
  \def\@bibitem#1{%
    \st@bibitem{#1}\a@@label{#1}\ignorespaces}
  \def\@lbibitem[#1]#2{%
    \st@lbibitem[#1]{#2}\a@@label{#2}\ignorespaces}
  \def\a@@label#1{%
    \gdef\a@lab{\smash{\normalfont\small#1}}
    \ifvmode
      \if@inlabel
        \global\setbox\@labels\hbox{%
          \llap{\a@lab\let\a@lab\relax
                \kern\@totalleftmargin\kern\marginparsep}%
          \box\@labels}%
      \fi
    \fi}
\fi

\documentclass[12pt,letterpaper]{article}

\usepackage{amsmath,amssymb,array,calc,rotating,epsfig,psfrag}
\usepackage[nosort]{cite}

\ifnum\draftcontrol=1 \tolerance=1000 \fi

\renewcommand\baselinestretch{1.25}
\renewcommand\section{\@startsection {section}{1}{\z@}%
                                   {-3.5ex \@plus -1ex \@minus -.2ex}%
                                   {2.3ex \@plus.2ex}%
                                   {\normalfont\large\bfseries}}
\renewcommand\subsection{\@startsection{subsection}{4}{\z@}%
                                   {-3.25ex\@plus -1ex \@minus -.2ex}%
                                   {1.5ex \@plus .2ex}%
                                   {\normalfont\normalsize\bfseries}}
\renewcommand\subsubsection{\@startsection{subsubsection}{3}{\z@}%
                                   {-3.25ex\@plus -1ex \@minus -.2ex}%
                                   {1.5ex \@plus .2ex}%
                                   {\normalfont\normalsize\it}}
\renewcommand\paragraph{\@startsection{paragraph}{4}{\z@}%
                                   {-3.25ex\@plus -1ex \@minus -.2ex}%
                                   {1.5ex \@plus .2ex}%
                                   {\normalfont\normalsize\bf}}
\renewcommand\subparagraph{\@startsection{subparagraph}{5}{\z@}%
                                   {-1.25ex\@plus -1ex \@minus -.2ex}%
                                   {0ex \@plus .2ex}%
                                   {\normalfont\normalsize\it}}

\def\ie{{\it i.e.}}
\def\eg{{\it e.g.}}

\def\revise#1       {\raisebox{-0em}{\rule{3pt}{1em}}%
                     \marginpar{\raisebox{.5em}{\vrule width3pt\
                     \vrule width0pt height 0pt depth0.5em
                     \hbox to 0cm{\hspace{0cm}{%
                     \parbox[t]{4em}{\raggedright\footnotesize{#1}}}\hss}}}}

\def\caln         {{\cal N}}
\def\calo         {{\cal O}}

\def\complex      {{\mathbb C}}

\def\projective   {{\mathbb P}}

\def\zet          {{\mathbb Z}}

\def\sqr#1#2{{\vcenter{\vbox{\hrule height.#2pt
 \hbox{\vrule width.#2pt height#1pt \kern#1pt
 \vrule width.#2pt}\hrule height.#2pt}}}}

\def\C3Z3{{\complex_3/\zet_3}}
\def\U{\it U}
\def\N{\caln}

\def\PI{$\hat{\text{\raisebox{.65em}{\rule{.95em}{.05em}}\hskip -1em PI}}$}

\catcode`\@=12

\begin{document}


\title{On Duality Walls in String Theory}

\pubnum{%
NSF-KITP-03-11 \\
hep-th/0301231}
\date{January 2003}

\author{Amihay Hanany$^{+}$ and Johannes Walcher$^{*}$ \\[0.4cm]
\it $^{+}$Center for Theoretical Physics \\
\it Massachusetts Institute of Technology \\
\it Cambridge, MA 02139, USA\\[0.2cm]
\it $^{*}$Kavli Institute for Theoretical Physics \\
\it University of California \\
\it Santa Barbara, CA 93106, USA \\[0.2cm]
}

\Abstract{Following the RG flow of an $\N=1$ quiver gauge theory
and applying Seiberg duality whenever necessary defines a duality
cascade, that in simple cases has been understood holographically.
It has been argued that in certain cases, the dualities will pile
up at a certain energy scale called the duality wall, accompanied
by a dramatic rise in the number of degrees of freedom. In string
theory, this phenomenon is expected to occur for branes at a
generic threefold singularity, for which the associated quiver has
Lorentzian signature. We here study sequences of Seiberg dualities
on branes at the $\C3Z3$ orbifold singularity. We use the naive
beta functions to define an (unphysical) scale along the cascade.
We determine, as a function of initial conditions, the scale of
the wall as well as the critical exponent governing the approach
to it. The position of the wall is piecewise linear, while the
exponent appears to be constant. We comment on the possible
implications of these results for physical walls. }


\makepapertitle

\body

\version\versionno

\section{Introduction}

Supersymmetric gauge theories are the most promising candidates
for grand unification in particle physics, and their dynamics is
therefore an important area of research. An even more interesting
subclass are gauge theories that can be embedded into string
theory, \ie, theories describing the low energy dynamics of
(decoupled subsectors of) appropriately compactified and branified
superstring theories. This subclass is distinguished by the fact
that string theory provides, in principle, the unification with
gravity.

At the basis of unification of course lie the phenomenon and
generalized notion of scale dependence of physical theories.
Effective gauge couplings and other physical parameters depend on
the energy scale at which they are measured, in a way that is
determined by renormalization group (RG) flow. Moreover, at
certain energy scales, even the elementary degrees of freedom can
change, leading to confinement and chiral symmetry breaking.
Prominent in supersymmetric gauge theories is the possibility of
duality. In particular, Seiberg duality \cite{seiberg} is the
statement that a collection of different $\N=1$ supersymmetric
gauge theories with gauge group and matter content related in a
particular way can provide different microscopic definitions of
one and the same underlying theory. Which microscopic description
is appropriate depends, again, on the energy scale.

We will here be concerned with one class of four-dimensional gauge
theories that can be embedded in string theory, namely through
D-branes at singularities of Calabi-Yau threefolds. For such
theories, some of the above questions have been entirely
reformulated in recent years in the light of the AdS/CFT
correspondence. For example, RG flow and duality can be realized
directly as the dependence of certain supergravity fields on the
portion of the dual geometry one is considering. Seiberg duality,
in particular, arises from the fact that a gauge theory
interpretation is possible only if the periods of supergravity
form fields lie in a certain range \cite{klst}. Using gauge
symmetries to shift the periods corresponds on the D-brane side to
a change of basis of fractional branes at the singularity, as
recently explained in \cite{bedo}.

In this paper, we consider quiver gauge theories arising from
D-branes at the $\C3Z3$ orbifold singularity. This is an interesting
example because it is among the simplest singularities that is
intrinsically three-dimensional (\ie, not related to an ADE
singularity on K3), and some of the important ingredients of $\N=1$
theories in four dimensions, such as chiral anomaly cancellation,
appear there for the first time. Of importance to us here is the
fact that the associated quiver is hyperbolic, \ie, with indefinite
Cartan matrix, whereas for simple singularities the quiver is
elliptic or at best parabolic. See \cite{fiol} for a definition and
discussion of these terms.

The goal underlying our study is to understand RG flows, their
cascades, and their walls, for quivers of generic threefold singularities.
Here, as in \cite{klst}, we refer as a duality cascade to the
sequence of Seiberg dualities that are necessary along the RG flow in
order to keep a gauge theory interpretation at every energy scale.
Such a cascade was studied in \cite{klst}, where the theory was the
affine $A_1$ quiver, dual to the conifold geometry. More such flows
were analyzed from a purely gauge theory point of view by Fiol
\cite{fiol}. It is pointed out in \cite{fiol} that for a generic
hyperbolic quiver, the scales at which one has to perform Seiberg
duality pile up at a certain finite energy scale called a duality
wall \cite{strassler}. The existence of this wall casts some doubt
at the possible UV completion of the theory.

All quivers whose cascades have been analyzed so far in
\cite{klst,fiol} are non-chiral, and either are elliptic or
parabolic, or else are hyperbolic but have no apparent embedding
into string theory, which might make the existence of the wall a
little less worrisome. Our example is probably the simplest that
is at the same time chiral, hyperbolic, and can be embedded
in string theory. The naive procedure that we will use illustrates
the behavior of duality cascades, and in particular the appearance
of walls, in such theories.

Our present results can, however, not be viewed as support for the
existence of duality walls in string theory.\footnote{We thank
Andreas Karch for a discussion on the role of anomalous
dimensions.} The only anomaly-free brane configuration at the
$\C3Z3$ orbifold is the regular D$3$-brane, and this theory is
conformal. After Seiberg duality, the naive beta functions do not
vanish anymore, and one might be led to the conclusion that one
can induce a cascade in this way. But a more careful analysis
involving the exact beta function shows that the Seiberg dual
theories are, in fact, also conformal and do not flow. In
contrast, our prescription involves following the {\it unphysical}
flow induced by the naive beta functions. This is the same
procedure that was also used in \cite{cfikv} for the same quiver
as ours, and in \cite{fiol} for other theories, for which it is
actually more justified. In \cite{cfikv}, the possible set of
Seiberg dualities was presented as a tree originating in the IR
with an infinite number of branch points as one proceeds to the
UV. In fact there is no RG flow along this tree. Using the exact
beta functions it is easy to see that all theories on this tree
are conformal. One can view the naive flows that we study here as
an organizing principle for trees of Seiberg dualities.

The question we shall ask is a simple thermodynamical one. Along
the flow, how does the number of degrees of freedom depend on
the scale? Obviously, this question only makes sense before the
wall, so we first have to know its position. As we shall see, the
position of the wall depends on the initial conditions that we specify.
For our example, this dependence is, surprisingly, simply piecewise
linear, see Fig.\ \ref{wall} in section \ref{general}. On the
other hand, the approach to the wall appears to be universal. The
critical exponent measuring the number of degrees of freedom is
one, independent of initial conditions.

As mentioned above, the appearance of walls is unphysical in our
example. This is the simplest example of a del Pezzo quiver, as
mentioned in \cite{cfikv}. It would be interesting to analyze the
RG flow behavior of the higher del Pezzos. Since there is more
freedom regarding anomaly cancellation, one can imagine that some
configurations will actually exhibit physical duality walls. On
the other hand, it might be true that walls are simply absent
generically in the UV behavior of gauge theories that appear in
string theory. We hope to return to this problem in the future.

We conclude this introduction with a few speculations concerning
the physical relevance of duality walls. If duality walls turn out
to exist in string theory, this raises a number of interesting
questions. For example, in a holographic picture, counting the number
of degrees of freedom is related to the entropy of black holes (branes).
See \cite{alex} for an exposition of this philosophy in the context
of the Klebanov-Strassler flow. If the number of degrees of freedom
diverges at a certain energy scale, the corresponding black holes
must be very peculiar. Moreover, the divergence of the number of
degrees of freedom for an observer probing at the scale of the wall
would imply the existence of highly mysterious singular points in the
closed string moduli space. This might also indicate the emergence of
new effective degrees of freedom as the UV completion of the theory.

Last but not least, we mention another, more mathematical, aspect
of our work. It is by now well-appreciated that quivers, their
algebras and their representations have an intrinsic connection to
geometry in the context of D-branes. see \eg,
\cite{ceva,zaslow,domo,dfr,hiv,berenstein} The representation theory of
quivers is a rather important but very hard, branch of
mathematics, see \eg, \cite{kac}. Just to mention one aspect, the
group of Seiberg duality transformations is the natural analog of
the Weyl group whose role in the representation theory of Lie
algebras is familiar. Understanding this group can be very
difficult. It is a natural question to ask whether the duality
cascades (subgroups of the duality group) induced by RG flow play
any particular role in the general representation theory of quivers.

\section{The underlying quiver}

\begin{figure}[h]
\begin{center}
\psfrag{n1}{$n_1$} \psfrag{n2}{$n_3$} \psfrag{n3}{$n_2$}
\psfrag{f1}{$f_1$} \psfrag{f2}{$f_2$} \psfrag{f3}{$f_3$}
\epsfig{file=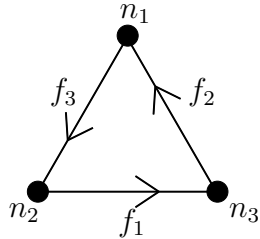,width=3cm} \caption{The quiver}
\label{quiver}
\end{center}
\end{figure}

In this paper, we study the properties of the tree of Seiberg dualities
of certain four-dimensional $\N=1$ supersymmetric gauge theories
embeddable in string theory through fractional branes at singularities.
Specifically, our gauge theory will be a quiver theory, with quiver
depicted in Fig.\ \ref{quiver}. The gauge group contains three factors
$\U(n_i)$, $i=1,2,3$, and there are $f_i$ chiral multiplets in bifundamentals
as shown.

There is also a superpotential, which is very important for
understanding the moduli space and the dynamics of the quiver.
In particular, it determines the anomalous dimensions of various
chiral fields. As mentioned in the introduction, we shall here
use naive beta functions without taking into account the effect of
the superpotential.

\subsection{D-brane interpretation}
\label{dbrane}

The gauge theory that canonically describes $N$ (regular)
D$3$-branes on the $\C3Z3$ orbifold is a quiver theory of the
above type with $(n_1,n_2,n_3)=(N,N,N)$ and
$(f_1,f_2,f_3)=(3,3,3)$ \cite{domo}. This has the D-brane
interpretation that the D$3$-brane can be obtained as a bound
state of the three fractional branes $e_i$ associated with each of
the nodes in Fig.\ \ref{quiver}, \ie, the $e_i$ correspond to the
gauge theories with $n_i=1$ and $n_j=0$ for $j\neq i$.

More generally, we can study arbitrary bound states
$(n_1,n_2,n_3)$ of the three fractional branes, with $f_i=3$ held
fixed. In fact, it is claimed that all (spacetime filling) branes
on the $\C3Z3$ orbifold can be constructed as bound states of the
elementary branes $(e_i)$, at least at zero string coupling. One
can also discuss, along similar lines, the spectrum of D-branes at
different points in the K\"ahler moduli space of the non-compact
Calabi-Yau $\calo_{\projective^2}(-3)$ into which the orbifold can
be blown up. We refer to \cite{dfr} for a general investigation of
the D-geometry of $\C3Z3$.

The point of interest here is that the $(n_1,n_2,n_3)=(N,N,N)$,
$f_i=3$ quiver is not the only possible theory whose moduli space
yields $\C3Z3$. Indeed, there is an infinite number of possible
duality transformations that one can apply to the above quiver
which give entirely equivalent descriptions. The D-brane
interpretation of this is that of a {\it change of basis of
elementary branes} \cite{bedo}. In other words, a given brane can
be constructed either as bound state of $(n_1,n_2,n_3)$ elementary
branes $e_i$ with $f_i$ chiral multiplets, or as bound state of
$(n_1',n_2',n_3')$ branes $e_i'$ with $f_i'$ chiral multiplets.
Since the Ramond-Ramond charge of the given brane does not depend
on the way we write it, we require that at the level of RR
charges, $\sum n_i e_i = \sum n_i' e_i'$. In other words, if we
have,
\begin{equation}
e_i' = \sum B_{ij} e_j \,, \eqlabel{BM}
\end{equation}
then
\begin{equation}
n_i' = \sum n_j B^{-1}_{ji} \,. \eqlabel{nodenm}
\end{equation}

To determine the number of chiral multiplets after the change of
basis, we note that we can assemble the $f_i$ into a $3\times 3$
matrix $I_{ij}$ which has the natural interpretation of
intersection form of cycles in the resolution
$\calo_{\projective^2}(-3)$ of the orbifold singularity $\C3Z3$,
see \eg, \cite{dfr}. As a consequence, we can read off the number
of chiral multiplets $f_i'$ from the intersection form in the new
basis, \ie,
\begin{equation}
I' = B I B^T \,. \eqlabel{Iprime}
\end{equation}

In the case at hand, as in many other examples, it turns out that
the total collection of quivers that give $\C3Z3$ can be
characterized by the solutions of a certain Diophantine equation,
which here is \cite{ceva,cfikv,fhhi}
\begin{equation}
n_1^2+n_2^2+n_3^2 = 3 n_1n_2n_3 \,. \eqlabel{diophantine}
\end{equation}
Namely, if $(n_1,n_2,n_3)$ is a solution of this equation with
${\rm g.c.d.} (n_1,n_2,n_3)=1$, then $N$ D$3$-branes on $\C3Z3$
are given by the quiver with charge vector $N(n_1,n_2,n_3)$ and
$f_i=3n_i$. In this way, each solution of \eqref{diophantine}
gives rise to a basis of fractional branes on $\C3Z3$. Of course,
it does not specify the superpotential, but as we noted before, we
shall neglect it here.

\subsection{Comments on RG flow}

Not every possible representation of the quiver gives rise to a
physical gauge theory at non-zero coupling. We also need to
satisfy the anomaly cancellation condition for every node (number
of incoming arrows equals number of outgoing arrows), \ie,
\begin{equation}
n_{i+1}f_{i-1} = n_{i-1} f_{i+1} \eqlabel{ancanc}
\end{equation}
for $i=1,2,3$ (we take $i=1,2,3\bmod 3$), in other words, we must
have $(f_1,f_2,f_3)\propto(n_1,n_2,n_3)$. Thus, there is only one
configuration for which anomalies are cancelled, which are exactly
the D$3$-branes. We shall therefore restrict ourselves to these
configurations.

The gauge theory on these D$3$-branes is a conformal theory. For
the canonical description, $(n_1,n_2,n_3)=(N,N,N)$, this follows
immediately from the naive beta functions, which read in general
\begin{equation}
\frac{d(1/g_i^2)}{d\ln\mu} = \beta_i = 3N_{c,i} - N_{f,i} \,,
\eqlabel{beta}
\end{equation}
where $N_{c,i}$ and $N_{f,i}$ are the number of colors and
flavors, respectively, on the $i$-th node.

Seiberg duality gives rise to equivalent theories and in
particular preserves the property of conformal invariance. To
compute the anomalous dimensions one can, for example, use the
Leigh-Strassler procedure.\footnote{We thank Andreas Karch for a
discussion on this point.} Denote the anomalous dimension of the
$i$-th fields by $\gamma_i$. These are related to the scaling
dimension by $D_i=1+{\frac12}\gamma_i$. Using the fact that the
superpotential is always cubic we have
$\gamma_1+\gamma_2+\gamma_3=0$. The numerator of the NSVZ beta
function then reads
\begin{equation}
\beta_i = 3n_i - 3n_{i-1}n_{i+1} + \frac{3}{2}
n_{i-1}n_{i+1}(\gamma_{i-1}+\gamma_{i+1}) = 3n_i
-3D_in_{i-1}n_{i+1}\,, \eqlabel{NSVZbeta}
\end{equation}
Equating this to zero gives the expression for the scaling
dimensions
\begin{equation}
D_i = \frac{n_i^2}{n_1n_2n_3}.
\end{equation}
As a check of this result we can verify that only solutions to the
Diophantine equation \eqref{diophantine} satisfy $D_1+D_2+D_3=3$
as required by the existence of such terms in the superpotential.

In all the computations which follow we will consider Seiberg
duality which applies only to cases in which the gauge group
factors are non-Abelian. For this reason we will need to have at
least 2 and in general $N$ D3 branes placed at the singularity.
However the value of $N$ does not play any crucial role in the
subsequent discussion and we will set it to $1$ for convenience of
the computation. We should have in mind that at any step of the
computation it is possible to restore $N$ to any desired value and
that any discussion about Seiberg duality which make sense apply
only to the SU part of the group and to $N\ge2$.

With this said, we will consider the gauge theory for a single
D$3$-brane, \ie, $(n_1,n_2,n_3)=(1,1,1)$, with respect to the
canonical basis of fractional branes. Applying Seiberg duality on
various nodes leads to other descriptions with different gauge
groups and matter content. All these theories are presumably
conformal, as explained above. In order to obtain a relation
between these various Seiberg dual theories, we introduce a
fictitious scale $\mu$, and let the couplings of the gauge theory
flow with $\mu$ according to the naive beta functions \eqref{beta},
applying Seiberg duality whenever one of the gauge couplings
diverges. In this way, we obtain a well-defined {\it duality
cascade}, that ends in the 'IR', $\mu\le\mu_0$, at the canonical
description $(N,N,N)$. Going back towards the 'UV', our
cascade then depends on the initial conditions at the scale $\mu_0$,
\ie, the gauge couplings, one of which must diverge. We imagine
$\mu_0$ being much smaller than the Planck scale.

We note that in the physical theory, the expected behavior is that
for any given initial conditions in any Seiberg dual quiver description
of the $(N,N,N)$ quiver, the theory will flow to the conformal fixed
point that is Seiberg dual to the original one. While this is difficult
to show in practice, the above arguments make it plausible.

\section{Duality cascades}

\subsection{Seiberg duality transformations}

Consider applying Seiberg duality to the gauge group on the $i$-th
node of the quiver. The number of colors and flavors for the $i$-th
node are given by $N_{c,i}= n_i$ and $N_{f,i}= n_{i+1} f_{i-1} =n_{i-1}
f_{i+1}$. Seiberg duality maps $N_{c,i} \mapsto N_{f,i}-N_{c,i}\,$.
Using the anomaly cancellation condition \eqref{ancanc}, this translates
into
\begin{align}
(n_{i-1},n_{i},n_{i+1})\mapsto (n_{i-1}',n_{i}',n_{i+1}')
&=(n_{i-1},3n_{i+1}n_{i-1} -n_i,n_{i+1})
\eqlabel{seiberg}\\
(f_{i-1}',f_{i}',f_{i+1}') &=3 (n_{i-1}',n_{i}',n_{i+1}')
\end{align}

Equation \eqref{seiberg} is really the simplest way of writing the
elementary duality transformation. However, to understand in
general what is going on at the level of D-branes, their charges,
and the relation to closed strings, it is necessary to keep in
mind that there are various other more powerful descriptions of
this duality. These formulations include Picard-Lefshetz
monodromy, toric duality, Weyl reflections, tilting equivalence of
derived categories, etc.. We will not go into full details here,
but just mention that the precise relation between these various
duality transformations is rather intricate and apparently not
fully understood at present.

\subsection{Coupling constants and energy scale}

We define $t=\ln \mu$ and parameterize the gauge coupling of the
$i$-th gauge group by
\begin{equation}
x_i=\frac{1}{g_i^2}  \,. \eqlabel{inversecoupling}
\end{equation}
We then have the RG flow equation
\eqref{beta}
\begin{equation}
\dot x_i = \frac{dx_i}{dt} = \beta_i = 3N_{c,i} - N_{f,i} \,,
\eqlabel{betap}
\end{equation}
where $N_{c,i}$ and $N_{f,i}$ are the number of colors and flavors
on the $i$-th node, respectively,
\begin{align}
N_{c,i} &= n_i \\
N_{f,i} &= n_{i+1} f_{i-1} = n_{i-1} f_{i+1} = 3 n_{i+1} n_{i-1} \,.
\end{align}
Thus,
\begin{equation}
\dot x_i = 3n_i - 3n_{i+1}n_{i-1} \,. \eqlabel{RG}
\end{equation}
It is instructive to check the beta function for the string
coupling $g_s$. In the canonical description of the $\C3Z3$
orbifold ($n_i=1$) the relation between the inverse gauge
couplings and the string coupling is given by
\begin{equation}
\frac{1}{g_s} = \sum_{i=1}^3\frac{1}{g_i^2} \,.
\end{equation}
This formula generalizes to the quiver of Fig.\ \ref{quiver} as
\begin{equation}
x\equiv\frac{1}{g_s} = \sum_{i=1}^3x_in_i\,.
\eqlabel{stringcoupling}
\end{equation}
{}From this and equation \eqref{RG} we can compute the
corresponding beta function for the string coupling
\begin{equation}
\dot x = 3(\sum_{i=1}^3n_i^2-3n_1n_2n_3)=0\,.
\eqlabel{Diophantine}
\end{equation}
The expression in brackets vanishes since it is precisely the
Diophantine equation \eqref{diophantine} for the $\C3Z3$ orbifold
as explained in detail in \cite{fhhi}. The result is that the
string coupling stays constant along the flow.

\subsection{The cascade}

As explained above, we start our (inverse) duality cascade at
$(n_i)=(1,1,1)$, $(f_i)=(3,3,3)$ by specifying three gauge
couplings, \ie, $x_1^{(0)},x_2^{(0)},x_3^{(0)}$, and applying
Seiberg duality to one of the nodes (without loss of generality,
let us say the third). We note immediately that this actually
requires $x_3^{(0)}=0$ at the scale $\mu_0$, so that we have one
less initial condition. At the first step, we then get
$(n_i)=(1,1,2)$, $(f_i)=(3,3,6)$ (henceforth, we shall suppress
the $f_i$). The theory starts flowing according to \eqref{RG} with
$t=\ln\mu$. We have
\begin{equation}
\dot x = (\beta_1,\beta_2,\beta_3) = (-3,-3,3)
\end{equation}
so that two couplings grow and one decreases. The next step in the
cascade happens when one of the two inverse couplings $x_1$, $x_2$
reaches zero. Which node we dualize on will depend in the initial
conditions. Let us assume that $x_1^{(0)}>x_2^{(0)}$. Then the
second dualization will happen after $\Delta t= x_2^{(0)}/3$,
a point at which $x_1=x_1^{(0)}- x_2^{(0)}$, $x_3=x_2^{(0)}$,
and $n=(1,5,2)$, and so on.

Before we proceed, we show that always two of the couplings grow
towards the UV, and one decreases \cite{cfikv}. Consider an
arbitrary step in the cascade, say we dualize on the second
node at $t=t_*$. For $t<t_*$, $\beta_2/3=n_2-n_1n_3<0$, and we
assume that one of $\beta_{1,3}$ is positive, and one negative.
After the duality, $t>t_*$, we have
\begin{align}
\beta_1'/3 &= n_1'-n_2'n_3' = n_3(n_2 - n_1n_3) + n_1(1-2n_3^2) < 0 \\
\beta_2'/3 &= n_2'-n_1'n_3' = 2n_1n_3- n_2 > 0 \\
\beta_3'/3 &= n_3'-n_1'n_2' = n_1(n_2 - n_1n_3) + n_3(1-2n_1^2) <
0 \,.
\end{align}
By induction, we see that at every stretch of the cascade, two
couplings grow towards the UV, and one decreases.

Our cascade is essentially a dynamical system given by \eqref{RG},
with the prescription to apply the duality \eqref{seiberg}
whenever any of $x_i=0$. One may view this dynamical system as a
``billiard'', in which $x_i=0$ behave like walls, with nonelastic
reflections at the walls (but note that also the position of the
walls of the billiard change after each reflection). The question
we would like to answer is how the $n_i$ behave as a function of
``time'' $t$, for given initial conditions $x_i^{(0)}$. One might
suspect this dependence to be rather sensitive, and the billiard
to display ``fractal'' behavior. We will see that this need not be
and that there are some quantities with rather simple behavior.

We make contact with \cite{cfikv} by noting that one can represent
all possible cascades by a ``tree'', in which at each node one
ingoing branch splits into two outgoing branches, encoding the
sequence of nodes of the quiver that one dualizes on. Our point
of view here is that this tree can be organized in a very efficient
and physically well motivated way by using the naive beta
functions. More precisely, the initial conditions select a
specific branch of the tree, making the system deterministic.

\section{One branch of the tree}
\label{branch}

Instead of being general, we now analyze one particular branch of
the above mentioned tree, in which only nodes $2$ and $3$
participate. We will see below that this can only be achieved by
choosing the singular initial condition $1/x_1^{(0)}=0$. To this
end, we will first give a slight reformulation of the dualities
that involves moving the nodes. Then we solve for this branch of
the tree explicitly.

\subsection{The cascade in terms of $(p,q)$ charges}
\label{pq}

One systematic way of describing the duality cascade is by using
the method of $(p,q)$ webs which was introduced in \cite{hi} and
was discussed in detail in \cite{fhhi,fh}. Let us review the
essential details needed for the present discussion.

Given a set of charges, $(p_i,q_i), i=1\ldots 3$, define the
intersection matrix $I_{ij}$ by
\begin{equation}
I_{ij} = p_iq_j-p_jq_i.
\end{equation}
This intersection matrix is an antisymmetric matrix with entries
that encode the quiver data,
\begin{equation}
f_i=\epsilon_{ijk}I_{jk}.
\end{equation}
This equation implies that the set of ranks of the gauge groups,
$n_i$, is a null vector of the intersection matrix $I_{ij}$ as
required by the anomaly cancellation condition \eqref{ancanc},
\begin{equation}
I_{ij}n_j = 0.
\end{equation}
In terms of $(p,q)$ charges, this anomaly cancellation condition
gets the form
\begin{align}
p_in_i&=0,
\eqlabel{anom}\\
q_in_i&=0.
\end{align}
To make connection with the RR charges $e$ discussed in section
\ref{dbrane} we note that setting \cite{hi}
\begin{align}
e_1 &=\frac{(p-q)^2-9(q^2-1)}{18q}, \\
e_2 &=\frac{p-q}{3}, \\
e_3 &=q,
\end{align}
provides a consistent set of the RR charges as expressed in
section \ref{dbrane}.

Seiberg duality on a single node in the quiver is a
Picard-Lefshetz monodromy on the $(p,q)$ charges. Suppose we start
by dualizing node 3. This is done by a monodromy action of the
$(p_2,q_2)$ charges on the charges $(p_3,q_3)$. Define $d$ to be
\begin{equation}
d = p_2 q_3 - p_3 q_2 \,.
\end{equation}
(We really have $d=3$, but the discussion for general $d$ is the
same.) Then the matrix $B$ in equation \eqref{BM} is
\begin{equation}
\eqlabel{B} B =\begin{pmatrix} 1 & 0 & 0 \\ 0 & 0 & -1 \\ 0 & 1 &
$d$
\end{pmatrix}\,.
\end{equation}
The new charges $(p',q')$ are given by the matrix action
\begin{align}
p'_i&=B_{ij}p_j,
\eqlabel{pqmono}\\
q'_i&=B_{ij}q_j.
\end{align}
The first row just keeps the charges $(p_1,q_1)$ as spectators,
not involved in the monodromy action. The second row replaces
position of the 3rd set of charges with the second. An additional
minus sign comes from the fact that the ranks, or node numbers,
$n_i$, change sign in order to satisfy the anomaly cancellation
condition, \eqref{anom}. The third row corresponds to the
Picard-Lefshetz monodromy. Note that the 2nd and 3rd gauge groups
have replaced their label. It is reassuring to find that using
these notations the new intersection matrix $I'$ is given in
equation \eqref{Iprime}, and the new node numbers are given in
equation \eqref{nodenm}. Furthermore, the new node numbers $n'$
and new charges $(p',q')$ satisfy the anomaly cancellation
conditions \eqref{anom}. It is easy to verify that the 23 matrix
entry is not changed, $I_{23}=I'_{23}$.

\subsection{Explicit solution}

The next step in the cascade that we are describing is given by
another application of the matrix $B$, \eqref{B} on the charges
$(p',q')$, and the full duality cascade simply becomes the successive
application of the matrix $B$. In order to solve explicitly this `one
branch of the tree' let us denote the node numbers $n_i$ after the
$k$-th step of the duality cascade by $n^{(k)}_i$. They are simply
obtained from the initial node numbers $n^{(0)}_i=1$ by the application
of the $k$-th power of the matrix $B$ to the initial charges $(p,q)$, \ie,
\begin{equation}
n^{(k)}_i = n^{(0)}_jB^{-k}_{ji}\,.
\end{equation}
By diagonalizing $B$, one can get an expression for these numbers in
terms of the eigenvalues $\lambda$ of the matrix $B$. These eigenvalues
are the solutions of the quadratic equation
\begin{equation}
\lambda^2-d\lambda+1=0\,, \qquad \lambda_{\pm}={\frac12}(d\pm
\sqrt{d^2-4})
\end{equation}
We then obtain for the node numbers
\begin{align}
n^{(k)}_1 &= 1, \eqlabel{n1} \\
n^{(k)}_2 &= n^{(k+1)}_3, \eqlabel{n2}\\
 n^{(k)}_3 &=
\frac{\lambda_+^k-\lambda_-^k+\lambda_-^{k-1}-\lambda_+^{k-1}}
{\lambda_+-\lambda_-}\,. \eqlabel{n3}
\end{align}
Before we proceed it is interesting to observe the behavior of the
cascade as a function of $d$. Clearly, $d=2$ is a critical number
which gives eigenvalues 1 and therefore produces a linear growth
of $n^{(k)}$ as a function of $k$. For $d\ge3$ there is an
exponential growth while for $d=1$ the eigenvalues are complex and
we expect some critical change in behavior. These phenomena do not
happen for the present case of study which is for $\C3Z3$. In more
complicated geometries we would expect to find some of this
interesting pattern to appear. In the terminology of \cite{fiol}
which is mentioned in the introduction, the cases $d=2,d>2,d<2$
correspond to an elliptic, hyperbolic and parabolic Cartan matrix,
respectively. Furthermore, the duality cascade studied in
\cite{klst} has $d=2$ and therefore is elliptic, corresponding to
a linear growth of the rank in the number of duality steps.

In order to trace the (unphysical) energy scale $\mu$ along the
cascade, we need to compute the naive beta functions at each step.
We denote the beta function of the $i$-th group between the $k$-th and
$k{+}1$-st step of the duality cascade, as in equation \eqref{betap},
by $\beta^{(k)}_i$, and similarly, the inverse gauge couplings of
equation \eqref{inversecoupling} by $x^{(k)}_i=\frac{1}{(g^{(k)}_i)^2}$.
The beta functions after the $k$-th step of the cascade are then
\begin{align}
\beta^{(k)}_1 &=3\bigl(1-n^{(k)}_2n^{(k)}_3\bigr)=
-3\bigl(n^{(k)}_2 -n^{(k)}_3\bigr)^2 < 0 , \\
\beta^{(k)}_2 &=3\bigl(n^{(k)}_2 - n^{(k)}_3\bigr)>0, \\
\beta^{(k)}_3 &=3\bigl(n^{(k)}_3 - n^{(k)}_2\bigr)=-\beta^{(k)}_2<0 .
\eqlabel{betas}
\end{align}
The second equality in the first line is made using the
Diophantine equation in \eqref{Diophantine}. It is easy to see
that $\beta_1$ and $\beta_3$ are negative while $\beta_2$ is
positive, independent of $k$. To write down the solution of the
equations $\dot x_i^{(k)}=\beta_i^{(k)}$, we define the energy scale
at which the $k$-th step of the duality cascade is performed by $t_k$.
The initial conditions for these equations are the inverse
gauge couplings $x^{(1)}_1(t_1)=x^{(0)}_1$, $x^{(1)}_2(t_1)=x^{(0)}_3$,
and $x^{(1)}_3(t_1)=x^{(0)}_2$, where $t_1=\ln\mu_0$ is the (arbitrary)
scale at which we start the cascade. Taking into account the permutation
of the 2nd and 3rd gauge groups at each step of the cascade, we find
for $t_{k}<t<t_{k+1}$
\begin{align}
\label{beta1}
x^{(k)}_1(t) &=\beta^{(k)}_1(t-t_{k})+x^{(k-1)}_1(t_{k}), \\
\label{beta2}
x^{(k)}_2(t) &=\beta^{(k)}_2(t-t_{k})+x^{(k-1)}_3(t_{k}), \\
x^{(k)}_3(t) &=\beta^{(k)}_3(t-t_{k})+x^{(k-1)}_2(t_{k}).
\end{align}
The $k{+}1$-st step of the cascade is done when one gauge coupling
diverges. The signs of the beta function then imply that it is
always the third gauge group which is dualized. The condition
becomes $x^{(k)}_3(t_{k+1})=0, k\ge1$. Taking this into account and
combining the second and third equation after setting $t=t_k$ we
find that, using \eqref{betas},
\begin{align}
\label{beta2con}
\beta^{(k)}_2(t_{k+1}-t_{k})&=\beta^{(k-1)}_2(t_{k}-t_{k-1})=\cdots
=\beta^{(2)}_2(t_3-t_2)\\
&=\beta^{(1)}_2(t_2-t_1)+x^{(0)}_3=x^{(0)}_2+x^{(0)}_3.
\end{align}
This difference equation then solves for the energy scale at the
$k{+}1$-st step
\begin{equation}
t_{k+1}=t_1-\frac{x^{(0)}_3}{\beta^{(1)}_2}+\bigl(x^{(0)}_2+
x^{(0)}_3\bigr)\sum_{j=1}^k\frac{1}{\beta^{(j)}_2}
\label{energyk}\end{equation} This result can be rewritten using
the explicit expressions for the beta functions in terms of the
eigenvalues of the monodromy matrix. We have
\begin{equation}
\beta_2^{(k)}= 3 \frac{\lambda_+^{k+1}-\lambda_-^{k+1}+
2\lambda_-^{k} - 2 \lambda_+^k - \lambda_-^{k-1}+\lambda_+^{k-1}}
{\lambda_+-\lambda_-} =
3(d-2)\frac{\lambda_+^k-\lambda_-^k}{\lambda_+-\lambda_-} \,,
\end{equation}
and hence
\begin{equation}
t_{k+1}=t_1-\frac{x^{(0)}_3}{3}+\bigl(x^{(0)}_2+x^{(0)}_3\bigr)
\frac{(\lambda_+-\lambda_-)}{3(d-2)}
\sum_{j=1}^k\frac{1}{\lambda_+^j-\lambda_-^j}\,. \eqlabel{tk}
\end{equation}
Let us study the values of the inverse gauge couplings at the
energy scales $t_k$. According to our boundary condition,
$x^{(k)}_3(t_{k+1})=0$. Using equations \eqref{beta2} and
\eqref{beta2con} we find that $x^{(k)}_2(t_{k+1})=x^{(0)}_2+
x^{(0)}_3$, independent of $k$. The most interesting result is
for $x^{(k)}_1(t_{k+1})$. We can either compute it directly using
equations \eqref{beta1} and \eqref{energyk}, or by observing that
the weighted sum of the couplings, equation \eqref{stringcoupling},
is constant along the flow. This value can simply be computed
for $k=1$ and we summarize the results below.
\begin{align}
x^{(k)}_1(t_{k+1}) &= x^{(0)}_1 -
(n^{(k)}_2-1)x^{(0)}_2-(n^{(k)}_2-2)x^{(0)}_3,
\eqlabel{x1} \\
x^{(k)}_2(t_{k+1}) &=x^{(0)}_2+x^{(0)}_3,
\eqlabel{x2} \\
x^{(k)}_3(t_{k+1}) &=0,
\eqlabel{x3} \\
\frac{1}{g_s} &= \sum_{i=1}^3n^{(k)}_ix^{(k)}_i =
x^{(0)}_1+2x^{(0)}_3+  x^{(0)}_2.
\end{align}
Since $n_2^{(k)}$ grows without bounds, this computation demonstrates
that the inverse coupling of the first gauge group, the one which is
not participating in the cascade along the specific branch that we
have been considering, reaches zero at some point, and does so in an
exponential fashion. Therefore, this cascade can not go on forever
without involving the first gauge group---unless we had actually
made $x_1^{(0)}=\infty$. In this case, which is a special case
of the ones considered in \cite{fiol}, we easily see from \eqref{tk}
and the fact that $\lambda_+>1$ for $d=3$, that $\lim_{k\to\infty}t_k
<\infty$. This is the simplest illustration of the wall phenomenon.

\section{The Wall}
\label{general}

The conclusion of the previous section forces us to consider
cascades that involve all three nodes. Unfortunately, the
combinatorics become quite involved, and we have not been
able to solve explicitly for the general cascade. But our
systems lends itself naturally to a numerical study, since, as
it turns out, the dualities converge exponentially towards the
wall.

\subsection{The position of the wall}

The (numerical) determination of the position of the wall becomes
extremely simple once we remove the redundancy from the initial
conditions. Recall that we were already forced to make
$x_3^{(0)}=0$. Moreover, it is easy to see from the homogeneity of
the equations \eqref{betap}, and from the explicit computation in
the previous section that we may rescale $x_2^{(0)}$ to one. From
the point of view of the dynamical system, this simply amounts to
a rescaling of ``time'' $t$. From a physical point of view,
this rescaling is not the choice of an energy scale, which is
$t_1=\ln\mu_0$, but rather amounts to a rescaling of the string
coupling $g_s$. In any case, this leaves us with a single initial
condition $x_1^{(0)}$. Finally, we note that we may also restrict
ourselves to $x_1^{(0)}>1$, since otherwise we simply exchange
$x_1^{(0)}$ and $x_2^{(0)}$.

With these initial conditions, the cascade proceeds as follows. As
long as $x_1>0$, we are on the `branch of the tree' described in
the previous section. When $x_1$ reaches zero between the $k$-th
and $k{+}1$-st step of the cascade, we have to start including the
first node in the cascade, and we do not know the general
solution.

\begin{figure}[ht]
\begin{center}
\psfrag{2}{$2$} \psfrag{4}{$4$} \psfrag{6}{$6$} \psfrag{8}{$8$}
\psfrag{10}{$10$} \psfrag{12}{$12$} \psfrag{14}{$14$}
\psfrag{0.1}{$0.1$} \psfrag{0.2}{$0.2$} \psfrag{0.3}{$0.3$}
\psfrag{0.4}{$0.4$} \psfrag{0.5}{$0.5$} \psfrag{x1}{$x_1^{(0)}$}
\psfrag{tw}{$\ln\Lambda_{\rm wall}$}
\epsfig{file=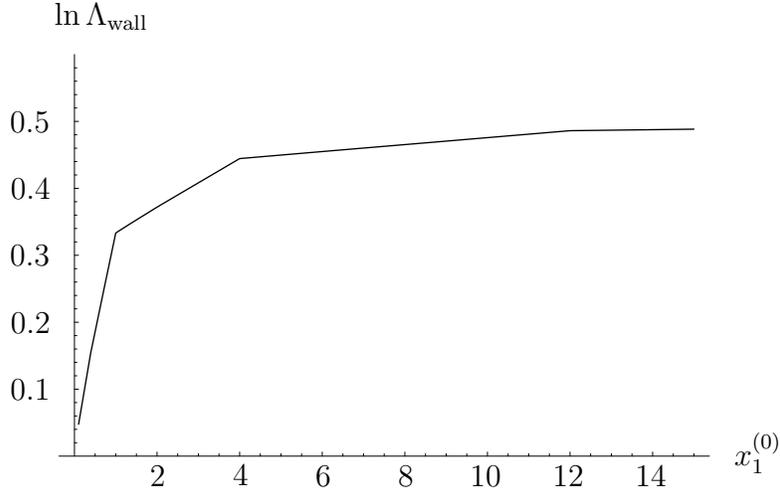,width=0.7\textwidth} \caption{The
position of the duality wall as a function of initial condition
$x_1^{(0)}$. The breaking points are located at
$x_1^{(0)}=1,4,12,33,\ldots$, and can be calculated from equation
\eqref{x1} as $n^{(k)}_2-1$. The asymptotic value
$\lim_{x_1^{(0)}\to\infty} t_{\rm wall}$ can be computed from
\eqref{tk} to be $.5117...$.} \label{wall}
\end{center}
\end{figure}

Our findings for the position of the wall $t_{\rm wall}=\ln
\Lambda_{\rm wall}$ as a function of the initial condition are
shown in Fig.\ \ref{wall}. Surprisingly, the function $t_{\rm
wall} (x_1^{(0)})$ is simply piecewise linear. The points of
discontinuity can be traced back to the explicit solution of the
 `branch of the tree' in the previous section. More precisely, it
appears that the cascade becomes singular for those initial
conditions for which the first node starts to play a role, \ie,
$x_1$ reaches $0$, at the same time $t_{k+1}$ at which we would
have had to perform the next step of dualizing node $3$. In other
words, the breaking points in Fig.\ \ref{wall} can be found by
setting $x_1^{(k)}(t_{k+1})$ to zero in \eqref{x1} and solving for
$x_1^{(0)}$. One then finds, using $x_2^{(0)}=1$ and
$x_3^{(0)}=0$, that the special initial conditions are given by
\begin{equation}
x_1^{(0),k} = n_2^{(k)} - 1 \,,
\end{equation}
where $n_2^{(k)}$ is given by equations \eqref{n2} and \eqref{n3}.
For $d=3$, this becomes the sequence $1,4,12,33\ldots$. (in Fig.\
\ref{wall}, there are also breaking points at the inverses of
these numbers, because of the symmetry $1\leftrightarrow 2$).
Moreover, we note that the position of the wall for these special
initial conditions $x_1^{(0),k}$ can also be determined from the
results of the previous section and are given by equation
\eqref{tk}.

At present, we do not understand the precise mechanism that leads
to discontinuities in $t_{\rm wall}(x_1^{(0)})$ at these special
points, nor the linear behavior between them. In fact, given the
origin of the $x_1^{(0),k}$ mentioned above, it would have been
natural to suspect that there be further breaking points whenever
two inverse couplings reach zero at the same time, possibly after
a fairly complicated sequence of dualities involving all three
nodes. We have not been able to detect such a `fractal' behavior.

\subsection{Critical exponent}

Given that we have found a simple description for the position of
the duality wall as a function of the initial condition
$x_1^{(0)}$, it is a natural question to ask how the wall is
approached, \ie, how do the node numbers $n_i$'s diverge as
$t\to t_{\rm wall}$. Most naively, one expects a power law
behavior
\begin{equation}
n_i(t) \sim \frac 1{(t_{\rm wall}-t)^{\gamma_i}} \,,
\eqlabel{gammai}
\end{equation}
and one can ask how the $\gamma_i$ depend on the initial
conditions. From a physical point of view, the `critical
exponents' $\gamma_i$ measure the growth in the number of degrees
of freedom during the approach of the wall. We can then also
reduce \eqref{gammai} to a single number, and study
\begin{equation}
n_1(t)\,n_2(t)\,n_3(t) \sim \frac 1{(t_{\rm wall}-t)^{\gamma}} \,,
\eqlabel{gamma}
\end{equation}
with $\gamma=\sum_{i=1}^3 \gamma_i$. It is clear that
\eqref{gamma} is not the only possible definition of $\gamma$, and
that one could imagine other measures of the `number of degrees of
freedom'. In a thermodynamical approach, however, the precise
definition should not matter too much.

As an example, let us look at the vicinity of the breaking points in
$t_{\rm wall}(x_1^{(0)})$. It is easy to see that, say for $k=2$, the
sequence of nodes that we dualize on is given by
\begin{align}
(321313131\ldots) &\qquad \text{for $x_1^{(0)}\lesssim x_1^{(0),2}$} \\
(323131313\ldots) &\qquad \text{for $x_1^{(0)}\gtrsim x_1^{(0),2}$} \,.
\end{align}
{}From this, one might be tempted to conclude that the second node is
not dualized on close to the wall, and $n_2$ is constant, which would
imply that $\gamma_2$ goes to zero at $x_1^{(0),2}$. However, as we
have seen in the previous section, as long as the initial conditions
are not singular (as for example at $x_1^{(0)}= x_1^{(0),2}$), the
cascade cannot proceed only with nodes $1$ and $3$. The second node
must eventually participate again.

In fact, it seems that rather than depending on initial conditions,
the critical exponents are actually constant. More precisely, the numerics
indicate that
\begin{equation}
- \frac{\ln n_1^{(k)} n_2^{(k)} n_3^{(k)}}{\ln(t_{\rm
wall}-t_k)} \longrightarrow 1 \qquad \text{as $k\to\infty$,}
\eqlabel{exponent}
\end{equation}
where $k$ numbers the steps in the general cascade starting at
initial condition $x_1^{(0)}$ and ending at $t_{\rm wall}$. We do
not have an analytical proof of \eqref{exponent}, but the following
heuristic arguments show that it is a sensible result.

We have seen that in the generic cascade, we must always include
all three nodes, and therefore all three node numbers should grow
in roughly equal proportions. It makes sense, therefore, to
consider appropriately averaged quantities\footnote{In the general
context of billiards and similar dynamical systems, one has to be
extremely careful with averagings of this sort. Here, they seem
to give sensible results.}, which we will denote by dropping the
subscript $i$. For example, in view of \eqref{gamma}, one could
consider the geometric average, \ie, $n^{(k)} = (n_1^{(k)}
n_2^{(k)}n_3^{(k)})^{1/3}$, etc.. Let us assume that in this
averaged sense, the node numbers grow exponentially with $k$, \ie,
\begin{equation}
\frac{\Delta n^{(k)}}{\Delta k} \sim n^{(k)} \,.
\eqlabel{heur1}
\end{equation}
Furthermore, we know from \eqref{stringcoupling} that the $x_i$
are inversely proportional to the $n_i$'s, \ie, $x^{(k)}\sim 1/n^{(k)}$,
while the beta functions grow quadratically in $n^{(k)}$. Hence,
\begin{equation}
\frac{\Delta t_k}{\Delta k}\sim x^{(k)} / \beta^{(k)}
\sim \bigl(n^{(k)}\bigr)^3 \,.
\eqlabel{heur2}
\end{equation}
Combining \eqref{heur1} and \eqref{heur2}, we find
\begin{equation}
\frac{\Delta n(t)}{\Delta t}\sim n(t)^4 \,,
\end{equation}
which indeed yields \eqref{exponent}, \ie, $n(t)\sim 1/(t_{\rm wall}-
t)^{1/3}$.

We note one caveat toward the result \eqref{exponent}, which comes
from the asymptotics along the special `branch of the tree'
described in the previous section. Indeed, from \eqref{n2},
\eqref{n3}, and \eqref{tk}, we find that for $k\to\infty$,
$t_\infty-t_k\sim 1/\lambda_+^k$, while $n_1^{(k)}=1$, and
$n_2^{(k)}, n_3^{(k)}\sim \lambda_+^{k}$. This would imply $\gamma
=2$. We attribute the discrepancy to the fact that
$x_1^{(0)}\to\infty$ is a singular initial condition.

\section{Conclusions and open questions}

In this paper, we have studied some simple properties of the
dynamical system \eqref{RG} associated by naive RG flow with the
quiver describing D$3$-branes at the $\C3Z3$ orbifold singularity.
The system exhibits the phenomenon of duality walls. The number
of degrees of freedom grows exponentially as a function of the
number of steps in the cascade. On the other hand, the growth in
scale decreases exponentially at the same rate as the number of
degrees of freedom. This results in the piling up of dualities at a
``duality wall'', \ie, the number of degrees of freedom grows faster
than exponential as a function of energy scale.

As mentioned in the introduction, our results do not have any
direct implications concerning the existence of duality walls
in string theory. These duality walls, introduced in
\cite{strassler} and studied in more detail in \cite{fiol},
thus await their realization in string theory. While one
could imagine that D$3$-branes at a generic threefold
singularity with a hyperbolic quiver would be the appropriate
framework for this, our present results are insufficient. Nevertheless,
our results give an illustration of the phenomenon, and at the
very least our flows can be viewed as an organizing principle
for the tree of Seiberg dualities.

We have in particular studied the dependence of the position of the
wall on the initial conditions of the cascade, and have found that
after the appropriate rescalings, the dependence is simply piecewise
linear. Moreover, the approach to the wall is found to be governed
by a simple scaling behavior \eqref{exponent}. While these results
are intriguing, in absence of a deeper understanding of the duality
walls, it is hard to give a physical interpretation for them.

We can imagine several approaches to the goal of answering some of
these physical questions. For instance, it would be useful to
understand the generic cascade analytically, generalizing our
results for the `branch of the tree' involving only two nodes. In
particular, one could try to verify more rigorously the results on
the position of the wall and the approach to it. In the quest for
further structure, and for a possible physical realization of the
walls, it will be helpful to repeat this analysis for other quivers
from branes at singularities, such as those arising from
contracting del Pezzos in Calabi-Yau threefolds.

More input can also be expected from looking for possible
holographic duals of our theory. Admittedly, since the
theory on regular D$3$-branes at the $\C3Z3$ orbifold singularity
is conformal, the holographic dual is simply ${\it AdS}_5\times
S^5/\zet_3$. Our cascades might be related to duals of irrelevant
deformations of such a background, and should make predictions, for
instance, about black hole entropy in this context.

Finally, we note that it would be interesting to look for the
special points in the K\"ahler moduli space corresponding to
the duality walls. Such points will be found because $x_1^{(0)}$
is the (appropriately rescaled) gauge coupling, hence related to
the B-field. What does the position of the wall $t_{\rm wall}
(x_1^{(0)})$ mean in K\"ahler moduli space?

\begin{acknowledgements}
We like to thank Alex Buchel, Sebastian Franco, Pavlos
Kazakopoulos, and Sonny Mantry for useful discussions. We would
like to thank Andreas Karch and David Tong for discussions on the
physical interpretation of our results. A.\ H.\ would like to
thank the Institute of Theoretical Physics at the Technion and the
Institute for Advanced Study at the Hebrew University for their
kind hospitality during various stages of this project. J.\ W.\
would like to thank the Perimeter \PI\ Institute for hospitality
during the final stages of this work. A.\ H.\ is partially
supported by the DOE under grant no.\ DE-FC02-94ER40818, by the
Reed Fund Award and by a DOE OJI award. The work of J.\ W.\ was
supported in part by the National Science Foundation under Grant
No.\ PHY99-07949.
\end{acknowledgements}


\begin{thebibliography}{99}

\bibitem{seiberg}
N.~Seiberg, ``Electric - magnetic duality in supersymmetric
nonAbelian gauge theories,'' Nucl.\ Phys.\ B {\bf 435}, 129 (1995)
[arXiv:hep-th/9411149].

\bibitem{domo}
M.~R.~Douglas and G.~W.~Moore, ``D-branes, Quivers, and ALE
Instantons,'' arXiv:hep-th/9603167.

\bibitem{klst}
I.~R.~Klebanov and M.~J.~Strassler, ``Supergravity and a confining
gauge theory: Duality cascades and $\chi SB$-resolution of naked
singularities,'' JHEP {\bf 0008}, 052 (2000)
[arXiv:hep-th/0007191].

\bibitem{alex}
A.~Buchel,
``Finite temperature resolution of the Klebanov-Tseytlin singularity,''
Nucl.\ Phys.\ B {\bf 600}, 219 (2001)
[arXiv:hep-th/0011146].

\bibitem{cfikv}
F.~Cachazo, B.~Fiol, K.~A.~Intriligator, S.~Katz and C.~Vafa, ``A
geometric unification of dualities,'' Nucl.\ Phys.\ B {\bf 628}, 3
(2002) [arXiv:hep-th/0110028].

\bibitem{fiol}
B.~Fiol, ``Duality cascades and duality walls,'' JHEP {\bf 0207},
058 (2002) [arXiv:hep-th/0205155].

\bibitem{strassler}
M.~J.~Strassler, ``Duality in Supersymmetric Field Theory and an
Application to Real Particle Physics,'' Talk given at
International Workshop on Perspectives of Strong Coupling Gauge
Theories (SCGT 96), Nagoya, Japan. Available at
http://www.eken.phys.nagoya-u.ac.jp/Scgt/proc/

\bibitem{bedo}
D.~Berenstein and M.~R.~Douglas, ``Seiberg duality for quiver
gauge theories,'' arXiv:hep-th/0207027.

\bibitem{dfr}
M.~R.~Douglas, B.~Fiol and C.~Romelsberger, ``The spectrum of BPS
branes on a noncompact Calabi-Yau,'' arXiv:hep-th/0003263.

\bibitem{hi}
A.~Hanany and A.~Iqbal, ``Quiver theories from D6-branes via
mirror symmetry,'' JHEP {\bf 0204}, 009 (2002)
[arXiv:hep-th/0108137].

\bibitem{fhhi}
B.~Feng, A.~Hanany, Y.~H.~He and A.~Iqbal, ``Quiver theories,
soliton spectra and Picard-Lefschetz transformations,''
arXiv:hep-th/0206152.

\bibitem{fh}
S.~Franco and A.~Hanany, ``Geometric dualities in 4d field
theories and their 5d interpretation,'' arXiv:hep-th/0207006.

\bibitem{ceva}
S.~Cecotti and C.~Vafa, ``On classification of N=2 supersymmetric
theories,'' Commun.\ Math.\ Phys.\  {\bf 158} (1993) 569
[arXiv:hep-th/9211097].

\bibitem{zaslow}
E.~Zaslow,
``Solitons and helices: The Search for a math physics bridge,''
Commun.\ Math.\ Phys.\  {\bf 175} (1996) 337
[arXiv:hep-th/9408133].

\bibitem{hiv}
K.~Hori, A.~Iqbal and C.~Vafa, ``D-branes and mirror symmetry,''
arXiv:hep-th/0005247.

\bibitem{berenstein}
D.~Berenstein, ``Reverse geometric engineering of singularities,''
JHEP {\bf 0204}, 052 (2002) [arXiv:hep-th/0201093].

\bibitem{kac}
V.~G.~Kac, ``Infinite root systems, representations of graphs and
invariant theory,'' Inv. Math. {\bf 56} (1980) 57.



\end{thebibliography}
\end{document}